\documentclass[prb,reprint,preprintnumbers,amsmath,amssymb,floatfix,aps,longbibliography]{revtex4-1}

\makeatletter
\def\@bibdataout@aps{%
 \immediate\write\@bibdataout{%
  @CONTROL{%
  apsrev41Control,author="08",editor="1",pages="0",title="0",year="1"%
  }%
 }%
 \if@filesw
  \immediate\write\@auxout{\string\citation{apsrev41Control}}%
 \fi
}%
\makeatother

\usepackage[utf8]{inputenc}
\usepackage[T1]{fontenc}
\usepackage{txfonts}
\usepackage{microtype}
\usepackage{eucal}
\usepackage{bm}
\usepackage{physics}
\usepackage{xr}
\usepackage{empheq}
\usepackage{lipsum}
\usepackage{subcaption}
\usepackage{color,verbatim}
\usepackage{soul}

\usepackage{graphicx}
\usepackage{float}
\usepackage[draft]{pgf}

\usepackage{booktabs}

\usepackage[colorlinks,allcolors=blue]{hyperref}
\usepackage[capitalize]{cleveref}

\usepackage{bibentry}
\newcommand{\ignore}[1]{}
\newcommand{\nobibentry}[1]{{\let\nocite\ignore\bibentry{#1}}}

\definecolor{red_lina}{rgb}{0.9,0,0}

\newcommand{\re}[1]{\mathrm{Re}\big\{#1\big\}}                  

\newcommand{\up}{\uparrow}                                      
\newcommand{\dn}{\downarrow}                                    

\definecolor{Haakon}{rgb}{0,0,1}

\newcommand{\ve}[1]{\boldsymbol{#1}}

\newcommand{\circcomm}[2]{
\left[ #1 \,\overset{\circ}{,}\; #2 \right]
}

\setcitestyle{numbers,square}

\begin{document}
\title{Spin pumping between noncollinear ferromagnetic insulators through thin superconductors}
\author{Haakon T. Simensen}
\author{Lina G. Johnsen}
\author{Jacob Linder}
\author{Arne Brataas}
\affiliation{Center for Quantum Spintronics, Department of Physics, Norwegian University of Science and Technology, NO-7491 Trondheim, Norway}

\date{\today}
 
\begin{abstract}
Dynamical magnets can pump spin currents into superconductors. To understand such a phenomenon, we develop a method utilizing the generalized Usadel equation to describe time-dependent situations in superconductors in contact with dynamical ferromagnets. Our proof-of-concept theory is valid when there is sufficient dephasing at finite temperatures, and when the ferromagnetic insulators are weakly polarized. We derive the effective equation of motion for the Keldysh Green's function focusing on a thin film superconductor sandwiched between two noncollinear ferromagnetic insulators of which one is dynamical. In turn, we compute the spin currents in the system as a function of the temperature and the magnetizations' relative orientations. When the induced Zeeman splitting is weak, we find that the spin accumulation in the superconducting state is smaller than in the normal states due to the lack of quasiparticle states inside the gap. This feature gives a lower backflow spin current from the superconductor as compared to a normal metal. Furthermore, in superconductors, we find that the ratio between the backflow spin current in the parallel and anti-parallel magnetization configuration depends strongly on temperature, in contrast to the constant ratio in normal metals. 
\end{abstract}

\maketitle


\section{Introduction}
Superconductivity and ferromagnetism are conventionally considered antagonistic phenomena. Superconductors (SCs) in contact with ferromagnets (FMs) lead to mutual suppression of both superconductivity and ferromagnetism \cite{Ginzburg1957, Bergeret2000}. Despite this apparent lack of compatibility, several intriguing effects also emerge from the interplay between superconductivity and ferromagnetism \cite{bergeret_rmp_05,Linder2015}. A singlet $s$-wave SC either in proximity with an inhomogeneous exchange field \cite{Bergeret2001}, or experiencing a homogeneous exchange field and spin-orbit coupling \cite{Gorkov2001,Bergeret2014}, induce spin-polarized triplet Cooper pairs. The generation of spin-polarized Cooper pairs is of particular interest, paving the way for realizing dissipationless spin transport \cite{Linder2015}. In recent developments, the combination of magnetization dynamics and superconductivity has gained attention. This is motivated by spin-pumping experiments reporting observations of pure spin supercurrents \cite{Jeon2018,Jeon2019}. Exhibiting a wide range of interesting effects and phenomena, SC-FM hybrids are promising material combinations in the emerging field of spintronics \cite{Eschrig2011}. 

It is well known that the precessing magnetization in FMs generates spin currents into neighboring materials via spin pumping \cite{Brataas2002,Tserkovnyak2002,Tserkovnyak2002a}. 
The injection of a spin current into a neighboring material generates a spin accumulation, which in turn gives rise to a backflow spin current into the FM. Spin pumping has a reactive and a dissipative component, characterized by how it affects the FM's dynamics. Reactive spin currents are polarized along the precession direction of the magnetization, $\dot{\ve{m}}$, and causes a shift in the FMR frequency. Dissipative spin currents resemble Gilbert damping and are polarized along $\ve{m} \times \dot{\ve{m}}$, relaxing the magnetization towards its principal axis. The dissipative spin current enhances the effective Gilbert damping coefficient \cite{Gilbert2004}, and broadens the FMR linewidth \cite{Tserkovnyak2002,Tserkovnyak2005}. 

In SCs, both quasiparticles and spin-polarized triplet Cooper pairs can carry spin currents. In the absence of spin-polarized triplet pairs, spin pumping is typically much weaker through a superconducting contact than a normal metal (NM) \cite{Bell2008, Morten2008}. The reduced efficiency is because the superconducting gap $\Delta$ prevents the excitation of quasiparticles by precession frequencies $\omega < 2\Delta$. When spin-polarized triplet pairs are present, spins can flow even for low FMR frequencies as pure spin supercurrents. Ref.~\cite{Jeon2018} reported evidence for such pure spin supercurrents. They measured an enhanced FMR linewidth in a FM-SC-heavy metal hybrid system as it entered the superconducting state, which is a signature of an enlarged dissipative spin current \footnote{\textit{Dissipative} here refers to its effect on the ferromagnet. In this sense, a dissipative spin current can still be carried through a SC without dissipation by spin-polarized triplet pairs.}. They attributed this observation to spin transport by spin-polarized triplet pairs. These findings and the rapid development of spintronics have lately sparked a renewed interest in spin transport through FM|SC interfaces \cite{Inoue2017,Jeon2018a,Yao2018,Taira2018,Montiel2018,Bobkova2018,Kato2019,Jeon2019,Golovchanskiy2019,Golovchanskiy2020} Several earlier works have also considered spin transport resulting from magnetization dynamics in SC-FM hybrids \cite{waintal_prb_02,houzet_prl_07, zhao_prb_08, konschelle_prl_09, yokoyama_prb_09,  teber_prb_10, linder_prb_11, kulagina_prb_14}.

Progress has been made in developing a theoretical understanding of the spin pumping through SCs \cite{Morten2008,Inoue2017,Yao2018,Taira2018,Montiel2018,Kato2019}. For instance, assuming suppression of the gap at the interface, Ref.~\cite{Morten2008} computed the reduced spin-pumping efficiency in the superconducting state using quasiclassical theory. However, to the best of our knowledge, a full understanding of the boundary conditions' complicated time-dependence between dynamical ferromagnets and superconductors is not yet in place. This development is required to give improved spin pumping predictions in multilayers of FMs, SCs and NMs. Furthermore, spin-pumping in superconducting systems with a noncollinear magnetization configuration remains theoretically underexplored, but can provide additional insight into the spin-transport properties. 

We present a self-consistent method designed to solve the explicit time-dependence arising from the magnetization dynamics by using the generalized Usadel equation. The explicit time-dependence complicates the treatment and understanding of the spin transport properties. We aim to describe a consistent proof-of-concept approach that is as simple as possible to understand. We will therefore use simplifying assumptions that are justified in weak insulating ferromagnets. Hopefully, the main message is then less hindered by subtleties. i) We explore trilayers with a thin film SC between two noncollinear FMIs. ii) We exclusively consider the imaginary part of the spin-mixing conductance in the contacts between the FMIs and the SC film. iii) We consider insulating ferromagnets. The first assumption requires that the interface resistance is larger than the superconductor's bulk resistance in the normal state, and that the superconductor is thinner than the coherence length. The second assumption is valid in weak ferromagnets. 

Our first main result is the equation of motion for the Green's function in the SC film when the magnetization precesses. 
Based on these results, we present quantitative predictions for the spin current as a function of temperature and the relative magnetization orientation between the FMIs. 


\section{The Generalized Usadel Equation and its Solution}
\begin{figure}[H]
\centering
\includegraphics[width=1\linewidth]{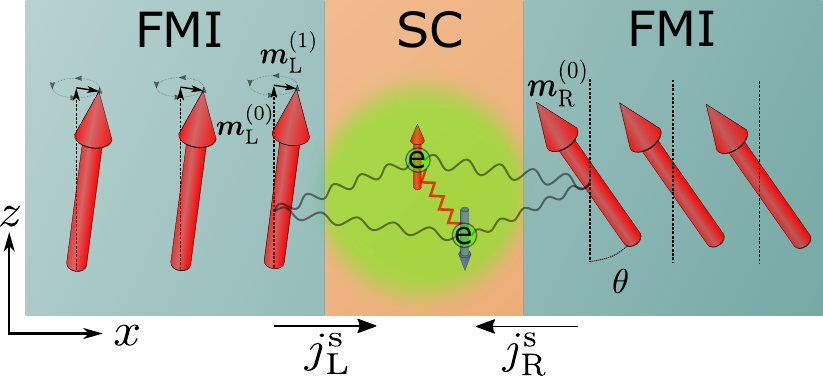}
\caption{FMI|SC|FMI trilayer. The superconductor is a thin film. The large red arrows depict the magnetic moments of localized $d$ electrons in the FMIs. The green cloud illustrates a gas of $s$ electrons with spin up (red) and down (blue). An attractive interaction between the $s$ electrons (red sawtooth-like line) gives rise to superconductivity. The $s$-$d$ exchange interaction at the interfaces gives rise to the indirect exchange interaction between the left and right FMI (wiggly grey lines). The precessing magnetization in the left FMI gives rise to spin currents $\ve{j}_{\rm L}^{\rm s}$ and $\ve{j}_{\rm R}^{\rm s}$ from the FMIs into the SC.}
\label{fig:schematic}
\end{figure}

In this section, we will first present the generalized Usadel equation taking into account the magnetization precession. We will demonstrate that it is possible to find an approximate solution to the time-dependence when the precession frequency is sufficiently slow. In superconductors, we will discuss how this approach requires sufficient dephasing since, otherwise, the peaks in the density of states invalidate the adiabatic assumption. Finally, we will solve the generalized Usadel equation and compute the resulting spin-current driven by the magnetization precession. Our analytical approach is supplemented by a numerical solution demonstrating the consistency of our assumptions.

\subsection{The Generalized Usadel equation in a FMI|SC|FMI trilayer}

The generalized Usadel equation determines the time evolution of the electron Green's function $\check{G}$ in the dirty limit. In a SC the generalized Usadel equation reads \cite{Usadel1970}
\begin{align}
\begin{split}
    -i D \nabla \check{G} \circ \nabla \check{G} + i \partial_{t_1}& \hat{\tau}_3 \check{G}(t_1,t_2) + i \check{G}(t_1,t_2) \partial_{t_2}  \hat{\tau}_3 \\
    &+ \circcomm{  \hat{\Delta}(t_1) \delta(t_1 - t')}{\check{G}(t',t_2)} = 0
    \label{eq:Usadel1},
\end{split}
\end{align}

\noindent where $D$ is the diffusion coefficient and $\delta(t)$ is the Dirac delta function. The symbol $\circ$ denotes time convolution, 
\begin{equation}
    {(a \circ b)(t_1, t_2) = \int_{-\infty}^{\infty} d t' a(t_1,t') b(t',t_2)},
\end{equation}
and ${\circcomm{a}{b} = a \circ b - b \circ a}$. $\check{G}$ and $\hat{\Delta}$ are matrices
\begin{align}
    \begin{split}
        \check{G} =
        \begin{pmatrix}
            \hat{G}^{\rm R} & \hat{G}^{\rm K} \\
            0 & \hat{G}^{\rm A}
        \end{pmatrix},
    \end{split}
    \begin{split}
        \hat{\Delta} = 
        \begin{pmatrix}
            0 & 0 & 0 & \Delta \\
            0 & 0 & -\Delta & 0 \\
            0 & \Delta^* & 0 & 0 \\
            -\Delta^* & 0 & 0 & 0
        \end{pmatrix},
    \end{split}
\end{align}

\noindent where R, A and K denote the retarded, advanced and Keldysh components, respectively. $\Delta$ is the superconducting gap. We choose to work in the gauge where $\Delta = \Delta^*$ is real. In our notation, the hat (e.g. $\hat{G}$) denotes $4 \times 4$ matrices in the subspace of particle-hole $\otimes$ spin space. The check (e.g. $\check{G}$) denotes matrices spanning Keldysh space as well. $\sigma_i$ are Pauli matrices spanning spin space, where $i \in \{0,\,x,\,y,\,z \}$ and $\sigma_0$ is the identity matrix. $\tau_i$ are Pauli matrices spanning particle-hole space, where $i \in \{0,\,1,\,2,\,3 \}$ and  $\tau_0$ is the identity matrix. 
To simplify the notation, we will omit outer product notation between matrices in spin and particle-hole space. Consequently, $\tau_i \sigma_j$ should be interpreted as the outer product of the matrices $\tau_i$ and $\sigma_j$. Moreover, we use the following notation for matrices that are identity matrices in spin space: $\hat{\tau}_i \equiv \tau_i \sigma_0$.
 
We consider thin film SCs sandwiched between two identical, homogeneous, weakly magnetized FMIs, illustrated in Fig.~\ref{fig:schematic}. Because of the insulating nature of the FMIs, we disregard any tunneling through the FMIs. The interaction between electrons in the SC region and the FMIs is therefore localized at the interfaces. This $s$-$d$ exchange interaction couples the localized $d$ electrons in the FMIs to the $s$ electrons in the SC at the interface. In thin film SCs, where the thickness of the superconductor is much shorter than the coherence length, $L_S \ll \xi_S$, we can approximate the effect of the $s$-$d$ exchange interaction as an induced, homogeneous magnetic field in the SC \cite{Hauser1969,Deutscher1969, Deutscher1969a,Hauser1971}. Furthermore, in computing the transport properties, this assumption requires that the interface resistances (inverse "mixing" conductances) are larger than the SC's bulk resistance in the normal state. When $L_S \ll \xi_S$, the Green's function changes little throughout the SC, and we therefore neglect the gradient term in the generalized Usadel equation within the SC. The resulting effective generalized Usadel equation for the FMI|SC|FMI trilayer then reads
\begin{align}
\begin{split}
    i \partial_{t_1} \hat{\tau}_3 \check{G}(t_1,t_2) + i &\check{G}(t_1,t_2) \partial_{t_2}  \hat{\tau}_3 + \circcomm{  \hat{\Delta}(t_1) \delta(t_1 - t')}{\check{G}(t',t_2)} \\
    &+ m_{\rm eff}\circcomm{\ve{m}(t_1)\cdot\boldsymbol{\hat{\sigma}} \delta(t_1 - t')}{\check{G}(t',t_2)}
    = 0,
    \label{eq:Usadel2}
\end{split}
\end{align}

\noindent where $\ve{m}(t) = \ve{m}_{\rm L}(t) + \ve{m}_{\rm R}(t)$, and where $\ve{m}_{\rm L/R}$ is the magnetization unit vector for the left/right FMI. $m_{\rm eff}$ is the effective magnetic field each of the two identical FMIs would separately induce in the SC (in units of energy), and $\hat{\boldsymbol{\sigma}} = \mathrm{diag}(\boldsymbol{\sigma},\boldsymbol{\sigma}^*)$, where $\boldsymbol{\sigma}$ is the vector of Pauli matrices in spin space. Note that when $\ve{m}_{\rm L} = -\ve{m}_{\rm R}$, the effective magnetic field in the superconductor vanishes, in agreement with the conclusions of Ref. \cite{DeGennes1966}.

The effective generalized Usadel equation \eqref{eq:Usadel2} was phenomenologically derived. We find the same equation by including boundary conditions to the FMIs \cite{cottet_prb_09, eschrig_njp_15}, and then averaging the Green's function over the thickness of the superconductor. 
In principle, one could also have included other terms which are higher order in both the Green's functions and magnetizations. However, we consider weak ferromagnets, where the phase difference $\Delta\varphi =\varphi_\up - \varphi_\dn$ in the spin-dependent reflection coefficients $r_{\up/\dn}$ is small. Then it is sufficient to include the imaginary part of the spin mixing conductance, which results in Eq.~\eqref{eq:Usadel2}. In other words, we disregard the real part of the mixing conductance, which is central in strong ferromagnets \cite{Morten2008}.


\subsection{Gradient expansion in time and energy}
\label{ssec:gradientexpansion}

The Green's function $\check{G}(t_1,t_2)$ correlates wave functions at times $t_1$ and $t_2$. By shifting variables to relative time ${\tau \equiv t_1 - t_2}$ and absolute time ${t \equiv (t_1 + t_2)/2}$, and performing a Fourier transformation in the relative time coordinate, the following identity holds \cite{Tikhonov2009, Rammer2012}
\begin{align}
    \mathcal{F}\left\{(a \circ b)(t_1,t_2)\right\} = \exp{\frac{i}{2}\left( \partial_E^a \partial_t^b - \partial_t^a \partial_E^b \right)} a(E,t) b(E,t),
    \label{eq:exp}
\end{align}

\noindent where $\mathcal{F}$ denotes Fourier transform in $\tau$, $a(E,t)$ and $b(E,t)$ are the Fourier transforms of $a(\tau, t)$ and $b(\tau, t)$ in the relative time coordinate, and $\partial^{a(b)}_{E(t)}$ denotes partial differentiation of the function $a$ ($b$) with respect to the variable $E$ ($t$). We will now Fourier transform and rewrite the generalized Usadel equation \eqref{eq:Usadel2} into $(E,\,t)$ coordinates.

The first two terms of Eq.~\eqref{eq:Usadel2} contain time differential operators. After rewriting these terms into the relative and absolute time coordinates, and Fourier transforming the relative time coordinate, we find \cite{Brinkman2003}
\begin{align}
    \begin{split}
        &\mathcal{F}\left\{ i \partial_{t_1} \hat{\tau}_3 \check{G}(t_1,t_2) + i \check{G}(t_1,t_2)  \partial_{t_2} \hat{\tau}_3 \right\} \\
        &= E \comm{\hat{\tau}_3}{\check{G}(E,t)} + \frac{i}{2} \acomm{\hat{\tau}_3}{\partial_t\check{G}(E,t)}.
    \end{split}
\end{align}

\noindent The remaining two terms in Eq.~\eqref{eq:Usadel2} contain commutators of time convolutions of one-point functions $\Delta(t_1)$ and $\ve{m}(t_1)$ and the Green's function $\check{G}(t_1,t_2)$. These two terms transform equally. We will, therefore, consider only the term containing the magnetization in detail. By straightforward substitution into the term containing the magnetization of Eq.~\eqref{eq:Usadel2} into Eq.~\eqref{eq:exp}, we find that
\begin{equation}
\begin{split}
&\mathcal{F}\left\{\circcomm{\ve{m}(t_1) \cdot \hat{\boldsymbol{\sigma}} \delta(t_1 - t')}{\check{G}(t',t_2)}\right\} \\
&= \exp{-\frac{i}{2} \partial_t^{\ve{m}} \partial_E^{\check{G}} } \ve{m}(t) \cdot \boldsymbol{\sigma} \check{G}(E,t) \\
&- \exp{\frac{i}{2} \partial_t^{\ve{m}} \partial_E^{\check{G}} } \check{G}(E,t) \ve{m}(t) \cdot \boldsymbol{\sigma} .
\end{split}
\end{equation}
In the following, we drop the arguments $E$ and $t$ to ease the notation.

We proceed by expanding the exponential function with differential operators,
\begin{align}
\begin{split}
    &\exp{-\frac{i}{2} \partial_t^{\ve{m}} \partial_E^{\check{G}} } (\ve{m} \cdot \boldsymbol{\sigma}) \check{G} - \exp{\frac{i}{2} \partial_t^{\ve{m}} \partial_E^{\check{G}} }  \check{G} (\ve{m} \cdot \boldsymbol{\sigma})\\
    &= \comm{(\ve{m} \cdot \hat{\boldsymbol{\sigma}})}{\check{G}} - \left(\frac{i}{2}\right) \acomm{\partial_t (\ve{m} \cdot \hat{\boldsymbol{\sigma}})}{\partial_E \check{G}} \\
    &+ \frac{1}{2!}\left(\frac{i}{2}\right)^2 \comm{\partial_t^2 (\ve{m} \cdot \hat{\boldsymbol{\sigma}})}{\partial_E^2 \check{G}} - \frac{1}{3!}\left(\frac{i}{2}\right)^3 \acomm{\partial_t^3 (\ve{m} \cdot \hat{\boldsymbol{\sigma}})}{\partial_E^3 \check{G}} + (...)
   \end{split}\label{eq:comm_expansion}
\end{align}

\noindent where $\acomm{...\,}{...}$ denotes an anticommutator. Here and further on, to ease the notation, we drop the superscript of the differential operators. Instead, we let the differential operators only act on the factor directly to the right of it. We keep terms only up to linear order in the gradients. This is justified when
\begin{align}
    \left\vert\frac{1}{2^3} \comm{\partial_t^n (\ve{m} \cdot \hat{\boldsymbol{\sigma}})}{\partial_E^n \check{G}}_{ij} \right\vert &\ll \left\vert\comm{\partial_t^{n-2} (\ve{m} \cdot \hat{\boldsymbol{\sigma}})}{\partial_E^{n-2} \check{G}}_{ij} \right\vert \label{eq:req1}, \\
    \left\vert\frac{1}{2^3} \acomm{\partial_t^n (\ve{m} \cdot \hat{\boldsymbol{\sigma}})}{\partial_E^n \check{G}}_{ij} \right\vert &\ll \left\vert\acomm{\partial_t^{n-2} (\ve{m} \cdot \hat{\boldsymbol{\sigma}})}{\partial_E^{n-2} \check{G}}_{ij} \right\vert \label{eq:req2},
\end{align}

\noindent where $\partial^n_t$ denotes the $n$-th partial derivative with respect to $t$. The magnetization precesses at at frequency $\omega$. Therefore, $\omega$ must be much smaller than the energy gradient of the Green's function. First, to avoid a diverging energy gradient of the Green's function, we assume finite temperatures. Second, we add a phenomenological dephasing parameter $\delta = 1/\tau_{\rm dep}$ to the Green's function, $E \rightarrow E + i \delta$, where $\tau_{\rm dep}$ is a characteristic dephasing time. We then find that the requirements \eqref{eq:req1}-\eqref{eq:req2} are satisfied when $(\omega \tau_{\rm dep})^2/8 \ll 1$ and $(\omega \beta)^2/8 \ll 1$, where $\beta = 1/k_{\rm B} T$ is the inverse temperature.

To linear order, the effective generalized Usadel equation in the FMI|SC|FMI trilayer reads
\begin{align}
    \begin{split}
        &E \comm{\hat{\tau}_3}{\check{G}} + \frac{i}{2}\acomm{\hat{\tau}_3}{\partial_t \check{G}} + \comm{\hat{\Delta}}{\check{G}} - \frac{i}{2} \acomm{\partial_t \hat{\Delta}}{\partial_E \check{G}} \\
        &\quad + m_{\rm eff} \comm{\ve{m} \cdot \hat{\boldsymbol{\sigma}}}{\check{G}} - \frac{i m_{\rm eff}}{2 } \acomm{\partial_t \ve{m} \cdot \hat{\boldsymbol{\sigma}}}{\partial_E \check{G}} = 0.
        \label{eq:Usadel3}
    \end{split}
\end{align}
In the next section, we will supplement this equation with terms arising from spin-memory loss.

\subsection{Spin relaxation}
\label{ssec:spinrelaxation}
To obtain a realistic model, we additionally need to include some sort of spin relaxation mechanism in the generalized Usadel equation \eqref{eq:Usadel3}. As a simple model, we model the relaxation as a coupling to a NM reservoir, parametrized by the coupling coefficient $V$. This coupling relaxes the Green's function in the SC towards the equilibrium solution around the Fermi level in the NM reservoir. The effective generalized Usadel equation including this relaxation reads
\begin{align}
    \begin{split}
        &E \comm{\hat{\tau}_3}{\check{G}} + \frac{i}{2}\acomm{\hat{\tau}_3}{\partial_t \check{G}} + \comm{\hat{\Delta}}{\check{G}} - \frac{i}{2} \acomm{\partial_t \hat{\Delta}}{\partial_E \check{G}} \\
        &\quad\quad\quad + m_{\rm eff} \comm{\ve{m} \cdot \hat{\boldsymbol{\sigma}}}{\check{G}} - \frac{i m_{\rm eff}}{2 } \acomm{\partial_t \ve{m} \cdot \hat{\boldsymbol{\sigma}}}{\partial_E \check{G}} \\
        &\quad\quad\quad\quad\quad\quad\quad\:\: + i V \comm{\check{N}}{\check{G}} - \frac{V}{2} \acomm{\partial_E \check{N}}{\partial_t \check{G}} = 0,
        \label{eq:Usadel4}
    \end{split}
\end{align}

\noindent where $\check{N}$ is the equilibrium Green's function in the NM reservoir. Additionally, this coupling gives a dephasing $E \rightarrow E - i V$ in the Green's function in the SC. This relaxation is therefore a possible source of the dephasing which we have already introduced in Sec.~\ref{ssec:gradientexpansion}.

\subsection{Parametrization}

We now aim to express the generalized Usadel equation~\eqref{eq:Usadel4} in a form that is easier to treat both analytically and numerically. We use a parametrization \cite{Ivanov2006} which maps the eight non-zero components of $\hat{G}^{\rm R}$, $\hat{G}^{\rm A}$ and $\hat{G}^{\rm K}$ onto two scalars (charge sector) and two vectors (spin sector), one of each reflecting the normal and anomalous parts of the Green's function. We expand the Green's function as (this applies to both the R, A and K components)
\begin{align}
    \hat{G} = \sum_{i \in \{ 0,1,2,3 \}} \sum_{j \in \{ 0,x,y,z \}} G_{i j} \tau_i \sigma_j,
    \label{eq:expansion}
\end{align}

\noindent where $G_{ij} = \frac{1}{4} \Tr{\hat{G} \tau_i \sigma_j}$. We gather the non-zero components into the following functions:
\begin{align}
\begin{split}
    G_0 &\equiv G_{30} , \\
    \ve{G} &\equiv [G_{0x}, G_{3y}, G_{0z}], \\
    F_0 &\equiv G_{1y}, \\
    \ve{F} &\equiv [-G_{2z}, G_{10}, G_{2x}].
\end{split}\label{eq:def1}
\end{align}

\noindent The scalar $G_0$ and the vector $\ve{G}$ describe the diagonal elements in particle-hole space of $\hat{G}$. The scalar $F_0$ and the vector $\ve{F}$ characterize the corresponding anomalous off-diagonal elements of $\hat{G}$. By inserting the definitions \eqref{eq:expansion}-\eqref{eq:def1} into the effective generalized Usadel equation~\eqref{eq:Usadel4}, we arrive at the following parametrized differential equations for the normal components,
\begin{widetext}
\begin{align}
        \frac{\partial G_0^{\rm R/A}}{\partial t} &= 
         m_{\rm eff} \left(\frac{\partial \ve{G}^{\rm R/A}}{\partial E}\right) \cdot \left(\frac{\partial \ve{m}}{\partial t}\right) - i
         \left(\frac{\partial F_0^{\rm R/A}}{\partial E}\right)\left(\frac{\partial \Delta}{\partial t}\right), \label{eq:Reffeq1} \\
        \frac{\partial \ve{G}^{\rm R/A}}{\partial t} &= 
        2 m_{\rm eff} \left( \ve{G}^{\rm R/A} \times \ve{m} \right) 
          +m_{\rm eff} \left(\frac{\partial G_0^{\rm R/A}}{\partial E}\right) \left(\frac{\partial \ve{m}}{\partial t}\right)
        -i \left(\frac{\partial \Delta}{\partial t}\right)\left( \frac{\partial \ve{F}^{\rm R/A}}{\partial E}\right), \label{eq:Reffeq2}\\
    \begin{split}
        \frac{\partial G_0^{\rm K}}{\partial t} &= 
         m_{\rm eff} \left(\frac{\partial \ve{G}^{\rm K}}{\partial E}\right) \cdot \left(\frac{\partial \ve{m}}{\partial t}\right) - i
         \left(\frac{\partial F_0^{\rm K}}{\partial E}\right)\left(\frac{\partial \Delta}{\partial t}\right)
         - 2 V \left[ G_0^{\rm K} - \left(G_0^{\rm R} - G_0^{\rm A}\right) \tanh\left(\frac{\beta E}{2}\right)  \right]  - i V \frac{\beta}{2}  \left( \frac{\partial G_0^{\rm R}}{\partial t} + \frac{\partial G_0^{\rm A}}{\partial t} \right) \sech^2\left(\frac{\beta E}{2}\right), 
    \end{split}\label{eq:Keffeq1} \\
    \begin{split}
        \frac{\partial \ve{G}^{\rm K}}{\partial t} &= 
        2 m_{\rm eff} \left( \ve{G}^{\rm K} \times \ve{m} \right) 
          +m_{\rm eff} \left(\frac{\partial G_0^{\rm K}}{\partial E}\right) \left(\frac{\partial \ve{m}}{\partial t}\right)
        -i \left(\frac{\partial \Delta}{\partial t}\right)\left( \frac{\partial \ve{F}^{\rm K}}{\partial E}\right)
        - 2 V \left[ \ve{G}^{\rm K} - \left(\ve{G}^{\rm R} - \ve{G}^{\rm A}\right) \tanh\left(\frac{\beta E}{2}\right)  \right]  - i V \frac{\beta}{2}  \left( \frac{\partial \ve{G}^{\rm R}}{\partial t} + \frac{\partial \ve{G}^{\rm A}}{\partial t} \right) \sech^2\left(\frac{\beta E}{2}\right).
    \end{split} \label{eq:Keffeq2}
\end{align}
\end{widetext}

\noindent We also obtain additional equations given in Appendix \ref{app:App1} for the anomalous components $F_0$ and $\ve{F}$ for both the R, A and K components. These equations \eqref{eq:AppReffeq3}-\eqref{eq:AppKeffeq4} are large and less transparent algebraic expressions. Last, we need the gap equation,
\begin{align}
\begin{split}
    \Delta &= - i \frac{N_0 \lambda}{4} \int_{- \omega_D}^{\omega_D}  \mathrm{d}E \: F_0^K,
\end{split}
\label{eq:gap}
\end{align}

\noindent where $\omega_D$ is the debye cut-off energy, $N_0$ is the Fermi-level electron density of states, and $\lambda$ is the BCS electron-phonon coupling constant. We will hereafter refer to $\Delta_0$ as the gap at zero temperature, and $\Delta$ as the gap at the temperature and effective magnetic field that is being considered.

For a self-consistent solution, all of the equations \eqref{eq:Reffeq1}-\eqref{eq:Keffeq2}, \eqref{eq:AppReffeq3}-\eqref{eq:AppKeffeq4} and \eqref{eq:gap}, are needed. If we assume a static gap however, only Eqs.~\eqref{eq:Reffeq1}-\eqref{eq:Keffeq2} are needed to determine the time evolution of the Green's functions once we know their solution at a given time $t$.

\subsection{Spin currents and effects on FMR}

The magnetization dynamics in FMs generate spin currents into neighboring materials. In the trilayer FMI|SC|FMI under consideration, these spin currents read
\begin{align}
    j_{X,x}^{\rm s} &= -\frac{i N_0 m_{\rm eff}}{8} \int_{-\infty}^{\infty} \mathrm{d}E \: \Tr{\sigma_x \tau_3  \circcomm{\ve{m}_X \cdot \hat{\boldsymbol{\sigma}}}{\check{G}}^K}, \label{eq:crntx} \\
    j_{X,y}^{\rm s} &= -\frac{i N_0 m_{\rm eff}}{8} \int_{-\infty}^{\infty} \mathrm{d}E \: \Tr{\sigma_y \tau_0  \circcomm{\ve{m}_X \cdot \hat{\boldsymbol{\sigma}}}{\check{G}}^K}, \label{eq:crnty} \\
    j_{X,z}^{\rm s} &= -\frac{i N_0 m_{\rm eff}}{8} \int_{-\infty}^{\infty} \mathrm{d}E \: \Tr{\sigma_z \tau_3  \circcomm{\ve{m}_X \cdot \hat{\boldsymbol{\sigma}}}{\check{G}}^K}, \label{eq:crntz}
\end{align}

\noindent where $\ve{m}_X$ is the magnetization at interface $X \in \{\mathrm{L},\,\mathrm{R} \}$, and where positive signs indicate spin-currents going from the FMIs into the SC. After expanding the convolution products in Eqs.~\eqref{eq:crntx}-\eqref{eq:crntz} to first order in time and energy gradients, we find
\begin{equation}
    \ve{j}_X^{\rm s} =  \frac{N_0 m_{\rm eff}}{4}\int_{-\infty}^{\infty} \mathrm{d}E \:\left[\frac{1}{2}\left(\frac{\partial G_0^{\rm K}}{\partial E}\right) \left(\frac{\partial \ve{m}_X}{\partial t}\right) + \left( \ve{G}^K \times \ve{m}_X \right)  \right].
          \label{eq:current}
\end{equation}

\noindent The first term in this expression is the so-called spin pumping current arising from the imaginary part of the mixing conductance. The spin pumping current equals ${\ve{j}^{\rm p} = (N_0 m_{\rm eff}/2) \partial \ve{m}_X/\partial t}$ both in SCs and NMs. The second term in Eq. \eqref{eq:current} is the backflow spin current $j^{\rm b}$ due to spin-accumulation in the SC \cite{Jiao2013}.
The spin pumping current is independent on temperature, relative magnetization angles, and on whether the system is superconducting or not. The backflow spin current depends on these system parameters, and will therefore be our main focus henceforth.

If we assume that the magnetizations of the FMIs are uniform, the Landau-Lifshitz-Gilbert equation for the left FMI can be written
\begin{align}
    \frac{\partial \ve{m}_{\rm L}}{\partial t} = -\gamma_0 \ve{m}_{\rm L} \times \ve{B}_{\rm eff} + \alpha_0 \left(\ve{m}_{\rm L} \times \frac{\partial \ve{m}_{\rm L}}{\partial t}\right) - \frac{\gamma_0}{M_s d} \ve{j}^{\rm s}_{\rm L},
\end{align}

\noindent where $\gamma_0$ is the gyromagnetic ratio of the ferromagnetic spins, $\ve{B}_{\rm eff}$ is the effective field in the FMI, $\alpha_0$ is the Gilbert damping parameter, $M_s$ is the saturation magnetization in the FMI, and $d$ is the thickness of the FMI. If we express $\ve{j}^{\rm s}_{\rm L}$ in reactive and dissipative components, $\ve{j}^{\rm s}_{\rm L} = C_r \frac{\partial \ve{m}_{\rm L}}{\partial t} + C_d \left(\ve{m}_{\rm L} \times \frac{\partial \ve{m}_{\rm L}}{\partial t}\right)$, we find the following renormalized properties in the FM:
\begin{align}
    \gamma_0 &\rightarrow \gamma = \frac{\gamma_0}{1 + \frac{C_r \gamma_0}{M_s d}}, \label{eq:gamma}\\
    \alpha_0 &\rightarrow \alpha = \frac{\gamma}{\gamma_0} \left( \alpha_0 + \frac{C_d \gamma_0}{M_s d} \right).\label{eq:alpha}
\end{align}
For later convenience, we define the reactive and dissipative spin currents, $\ve{j}^{\rm s}_{\rm r} \equiv C_r \frac{\partial \ve{m}_{\rm L}}{\partial t}$ and $\ve{j}^{\rm s}_{\rm d} \equiv C_d \left(\ve{m}_{\rm L} \times \frac{\partial \ve{m}_{\rm L}}{\partial t}\right)$.

\section{Results and discussions}

We will now use the equations of motion of \eqref{eq:Reffeq1}-\eqref{eq:Keffeq2}, \eqref{eq:AppReffeq3}-\eqref{eq:AppKeffeq4}, and the gap equation \eqref{eq:gap}, to find the spin current generated by FMR in a FMI|SC|FMI trilayer. We consider homogeneous magnetizations $\ve{m}_{\rm L}$ and $\ve{m}_{\rm R}$ in the left and right FMIs, respectively. The angle between the principal axes of the magnetizations is $\theta$. The left magnetization is precessing circularly around its principal axis at a precession angle $\varphi$ with angular frequency $\omega$. The right magnetization is static. The system is illustrated in Fig.~\ref{fig:schematic}. 

We will initially search for an analytical solution by treating the dynamic magnetization component as a perturbation from an equilibrium solution. Due to the complexity of the equations, we first assume that the gap is static. This approximation enables us to solve the problem for arbitrary relaxation $V$. Sec. \ref{ssec:lin1} presents this analytical approach. In principle, it is also possible to find a self-consistent analytical solution. However, the solution becomes extremely complex in the presence of relaxation due to the coupling between the retarded/advanced and Keldysh Green's functions. Hence, the full self-consistent problem is better suited for numerical treatments. In Sec. \ref{ssec:num}, we compare the results of a self-consistent numerical solution to the analytical solution in Sec. \ref{ssec:lin1}. We additionally outline a self-consistent analytical solution in App. \ref{ssec:lin2} in the absence of relaxation. This latter solution has restricted physical relevance, but is supplied for the convenience of further work in this framework.

\subsection{Analytical solution with static gap approximation}
\label{ssec:lin1}

We first separate the magnetization vector $\ve{m}$ into a static and a dynamic component, $\ve{m} = \ve{m}^{(0)} + \ve{m}^{(1)}$. The static component $\ve{m}^{(0)} = \ve{m}^{(0)}_{\rm L} + \ve{m}^{(0)}_{\rm R}$ is the sum of the static magnetizations of the left and right FMIs. The dynamic component $\ve{m}^{(1)}$ is the dynamic part of $\ve{m}_{\rm L}$. It has magnitude $\delta m$ and precesses around the $z$ axis with angular frequency $\omega$, $\ve{m}^{(1)} = \delta m \left[\cos(\omega t),\,\sin(\omega t),\,0 \right]$. This decomposition of the magnetization vectors is illustrated in Fig.~\ref{fig:schematic}. We now assume that: i) The dynamic magnetization component is much smaller than the gap, $m_{\rm eff} \delta m \ll \Delta$. ii) The fluctuations in the gap are much smaller than the dynamic magnetization amplitude, $\delta \Delta \ll m_{\rm eff} \delta m$.

Assumption i) enables us to expand the Keldysh Green's function components in the perturbation $\delta m$,
\begin{align}
\begin{split}
    G_0^{\rm K} &= G_0^{\rm K(0)} + G_0^{\rm K(1)} + (...),  \\
    \ve{G}^{\rm K} &= \ve{G}^{\rm K(0)} + \ve{G}^{\rm K(1)} + (...),
 \end{split}\label{eq:pert1}
\end{align}

\noindent where the $n$'th order terms are assumed to be $\propto \delta m^n$. We consider the first order expansion in $\delta m$ only, and choose therefore to disregard 2nd and higher order terms. Assumption ii) implies that the generalized Usadel equations for the advanced and retarded Green's functions [Eqs. \eqref{eq:Reffeq1}-\eqref{eq:Reffeq2}] decouple from the Keldysh component. In what follows, we will derive the solution for the Keldysh component. The retarded/advanced Green's functions can then be found simply by substituting K$\rightarrow$R/A and by setting $V = 0$ in the Keldysh component solution. 

To first order in $\delta m$, the effective generalized Usadel equations for the Keldysh component reads
\begin{align}
    \begin{split}
        \frac{\partial G_0^{\rm K (1)}}{\partial t} &= 
         m_{\rm eff} \left(\frac{\partial \ve{G}^{\rm K (0)}}{\partial E}\right) \cdot \left(\frac{\partial \ve{m}^{(1)}}{\partial t}\right) \\
         &- 2 V \left[ G_0^{\rm K(1)} - \left( G_0^{\rm R(1)} - G_0^{\rm A(1)} \right) \tanh\left(\frac{\beta E}{2}\right) \right], \label{eq:effeq21}
    \end{split}\\
    \begin{split}
        \frac{\partial \ve{G}^{\rm K (1)}}{\partial t} &= 
        2 m_{\rm eff} \left( \ve{G}^{\rm K (0)} \times \ve{m}^{(1)} + \ve{G}^{\rm K (1)} \times \ve{m}^{(0)} \right)\\
        &+ m_{\rm eff} \left(\frac{\partial G_0^{\rm K (0)}}{\partial E}\right) \left(\frac{\partial \ve{m}^{(1)}}{\partial t}\right) \\
        &- 2 V\left[ \ve{G}^{\rm K(1)} - \left( \ve{G}^{\rm R(1)} - \ve{G}^{\rm A(1)} \right) \tanh\left(\frac{\beta E}{2}\right)\right]. \label{eq:effeq22}
    \end{split}
\end{align}

\noindent We propose the ansätze
\begin{align}
\begin{split}
    G_0^{\rm K (1)} &= G_{0 +}^{\rm K (1)} e^{i \omega t} + G_{0 -}^{\rm K (1)} e^{-i \omega t}, \\
    \ve{G}^{\rm K (1)} &= \ve{G}^{\rm K (1)}_{+} e^{i \omega t} + \ve{G}^{\rm K (1)}_{-} e^{-i \omega t}.
\end{split}\label{eq:ans1}
\end{align}
After inserting the ansätze in Eq.~\eqref{eq:ans1} into Eqs.~\eqref{eq:effeq21}-\eqref{eq:effeq22}, we note that the differential equations separate into decoupled equations for the $+/-$ components. By solving for $G_{0 \pm}^{\rm K (1)}$ and $\ve{G}_{\pm}^{\rm K (1)}$, we obtain
\begin{align}
\begin{split}
    G_{0 \pm}^{\rm K(1)} &=  m_{\rm eff}\frac{ \pm\omega }{\pm\omega + 2 V} \left(\frac{\partial \ve{G}^{(0)}}{\partial E}\right) \cdot \ve{m}_{\pm}^{(1)} \\
    &\quad + \frac{2 V}{\pm\omega + 2V} \left( G_0^{\rm R(1)} - G_0^{\rm A(1)} \right) \tanh\left(\frac{\beta E}{2}\right), \label{eq:G01}
    \end{split} \\
    \ve{G}_{\pm}^{\rm K(1)} &= m_{\rm eff} \mathbf{A}_{\pm\omega}^{-1} \mathbf{B}_{\pm\omega} \ve{m}^{(1)}_{\pm} 
    + 2 V \mathbf{A}_{\pm\omega}^{-1} \left( \ve{G}^{\rm R(1)} - \ve{G}^{\rm A(1)} \right) \tanh\left(\frac{\beta E}{2}\right)
    \label{eq:G1},
\end{align}

\noindent where the matrices $\mathbf{A}_{\pm\omega}$ and $\mathbf{B}_{\pm\omega}$ are defined as
\begin{align}
    \begin{split}
        \mathbf{A}_{\pm\omega} =
        \begin{pmatrix}
            \pm i \omega + 2V & -2 m_{\rm eff} m_z^{(0)} & 2 m_{\rm eff} m_y^{(0)} \\
            2 m_{\rm eff} m_z^{(0)} & \pm i \omega + 2V & -2 m_{\rm eff} m_x^{(0)} \\
            -2 m_{\rm eff} m_y^{(0)} & 2 m_{\rm eff} m_x^{(0)} & \pm i \omega + 2V
        \end{pmatrix},
    \end{split}\\
    \begin{split}
        \mathbf{B}_{\pm\omega} =
        \begin{pmatrix}
            \pm i \omega C(E)  & -2 G_z^{\rm K(0)} & 2 G_y^{\rm K(0)} \\
            2 G_z^{\rm K(0)} & \pm i \omega C(E)  & -2 G_x^{\rm K(0)} \\
            -2 G_y^{\rm (0)} & 2 G_x^{\rm K(0)} & \pm i \omega C(E)
        \end{pmatrix},
    \end{split}
\end{align}
and where 
\begin{align}
\begin{split}
    C&(E) = \tanh\left(\frac{\beta E}{2}\right) \left(\frac{\partial G_0^{\rm R(0)}}{\partial E} - \frac{\partial G_0^{\rm A(0)}}{\partial E} \right) \\
    &+ \left( \left[G_0^{\rm R(0)} - G_0^{\rm A(0)}\right] - i V \left[ \frac{\partial G_0^{\rm R(0)}}{\partial E} + \frac{\partial G_0^{\rm A(0)}}{\partial E} \right] \right) \frac{\beta}{2} \sech^2\left(\frac{\beta E}{2}\right).
    \label{eq:C}
\end{split}
\end{align}

The solution to $\ve{G}^{\rm K(1)}$ is particularly simple when $\theta = 0$ or $\theta =\pi$. For $\theta = 0$, we obtain
\begin{align}
\begin{split}
    &\ve{G}^{\rm K(1)}_{\theta = 0} = m_{\rm eff} \left( 2 G_z^{\rm K(0)} +  \omega C(E) \right) \frac{(4 m_{\rm eff} + \omega) \ve{m}^{(1)} + \frac{2 V}{\omega}  \frac{\partial \ve{m}^{(1)}}{\partial t}}{(2V)^2 + (4 m_{\rm eff} + \omega)^2} \\
    &+  \frac{2 V^2 \ve{m}^{(1)} - \frac{V}{\omega} (4 m_{\rm eff} + \omega) \frac{\partial \ve{m}^{(1)}}{\partial t} }{(2V)^2 + (4 m_{\rm eff} + \omega)^2} G_z^{\rm K(0)}, 
    \label{eq:analytical1}
\end{split}
\end{align}
where we have inserted $m_z^{(0)} = 2$. We observe that a finite $V$ introduces a component of $\ve{G}^{\rm K}$ parallel to $\partial \ve{m}_{\rm L}/\partial t$. When we insert this component into the spin current in Eq.~\eqref{eq:current}, we see that it generates both a reactive and a dissipative backflow current, $\ve{j}^{\rm b}_{\rm r}$ and $\ve{j}^{\rm b}_{\rm d}$. Hence, even though the spin pumping current is purely reactive, the backflow spin current can indeed carry a dissipative part due to relaxation in the SC. Moreover, we note that the effective magnetic field $2 m_{\rm eff}$ suppresses the amplitude of $\ve{G}^{\rm K(1)}$. This feature is due to Hanle precession of $\ve{G}^{\rm K(1)}$ around the effective magnetic field, which reduces the effect of the excitation. 

When $\theta = \pi$, the Hanle precession is more or less absent due to a very small effective magnetic field $\propto m_{\rm eff} \sin\varphi$. Under the assumption that the precession angle is sufficiently small, $\sin\varphi \ll \omega/m_{\rm eff}$, we obtain
\begin{align}
\begin{split}
    &\ve{G}^{\rm K(1)}_{\theta = \pi} = m_{\rm eff} \omega C(E)  \frac{\omega \ve{m}^{(1)} + \frac{2 V}{\omega}  \frac{\partial \ve{m}^{(1)}}{\partial t}}{(2V)^2 + \omega^2}, 
    \label{eq:analytical2}
\end{split}
\end{align}
\noindent As a control check, we can verify that we obtain the instantaneous equilibrium solution $(\ve{G}^{\rm R(1)} - \ve{G}^{\rm A(1)})\tanh(\beta E/2)$ when $V \gg \omega$.

In the second line of $C(E)$ in Eq.~\eqref{eq:C}, we have isolated the source of non-equilibrium behavior of $\ve{G}^{\rm K}$. This non-equilibrium part arises from the energy gradient of the distribution function, and is therefore proportional to $\sech^2\left(\frac{\beta E}{2}\right)$. In the normal metal limit, we have $\partial G_0^{\rm R}/\partial E = 0$, and ${\int_{-\infty}^{\infty} \mathrm{d}E \; C(E) = 4}$ is therefore constant and independent of temperature. The spin current is therefore independent of temperature in the NM limit.

The coefficient $C(E)$ in Eq.~\eqref{eq:C} predicts that the non-equilibrium effects mostly arise within a thermal energy interval $\pm \beta^{-1}$ from the Fermi level. There are two tunable parameters that affect the number of quasiparticle states within this energy interval in a SC: First, at higher temperatures, the energy interval in which quasiparticles can be excited broadens. The more overlap there is between this energy window and the gap edge, the larger we expect the spin accumulation to be. Another thermal effect is that the gap $\Delta$ decreases with increasing temperature, which enhances the abovementioned effect. Second, the effective magnetic field introduces a spin-split density of states, which pushes half of the quasiparticle states closer to the Fermi level. An additional effect is that the gap decreases with an increasing effective magnetic field, an effect which moreover is temperature dependent. Therefore, the effective magnetic field also affects the number of quasiparticle states within a thermal energy interval from the Fermi level. Both the temperature and effective magnetic field can hence be tuned to increase the spin accumulation. The spin accumulation in turn generates a backflow spin current into the FMIs. We therefore expect a larger backflow spin current from a SC at higher temperatures and for stronger effective magnetic fields.

We will now evaluate the angular and temperature dependence of the backflow spin current for a particular FMI|SC|FMI trilayer. We choose the parameters in the SC so that they match those of Nb. That is, we choose ${1/V = \tau_{\rm sf}/2\pi \sim 10^{-10}}$ s \cite{Johnson1994} and a critical temperature ${T_{\rm c} = 9.26}$ K \cite{Seidel1958}. Moreover, we use an effective magnetic field strength ${m_{\rm eff} = 0.1 \Delta_0}$, and a magnetization precession angle ${\varphi = \arcsin(0.01)}$. Last, we use a precession frequency ${\omega = 0.005 \Delta_0 \approx 10 \;\mathrm{GHz}}$, which is an appropriate frequency for e.g. yttrium iron garnet (YIG). The relaxation introduces a dephasing ${V = 0.05 \Delta_0}$, which is sufficient to justify the gradient expansion. The gap ${\Delta = \Delta(T,\theta,m_{\rm eff})}$ is found by solving the gap equation self-consistently \cite{Zheng2000} to zeroth order in the dynamic magnetization, as well as checking that the free energy of the superconducting state is lower than in the normal metal state. The assumptions i) and ii) underlying the static gap approximation can be satisfied for any effective field $m_{\rm eff}$ providing we choose an appropriate precession amplitude, $\delta m$, which can be tuned with the AC magnetic field used to excite FMR in the FMI.

In the FMI|SC|FMI trilayer, the expression for the backflow spin current in Eq. \eqref{eq:current} implies that there is a static RKKY contribution to the spin current. This RKKY contribution is due to the finite $\ve{G}^{\rm K}$ close to the Fermi level. However, other terms also contribute to the RKKY interaction beyond the quasiclassical theory. Therefore, we subtract the instantaneous RKKY-like static contribution to the spin current.
\begin{figure}[htb]
\centering
\includegraphics[width=1.0\columnwidth]{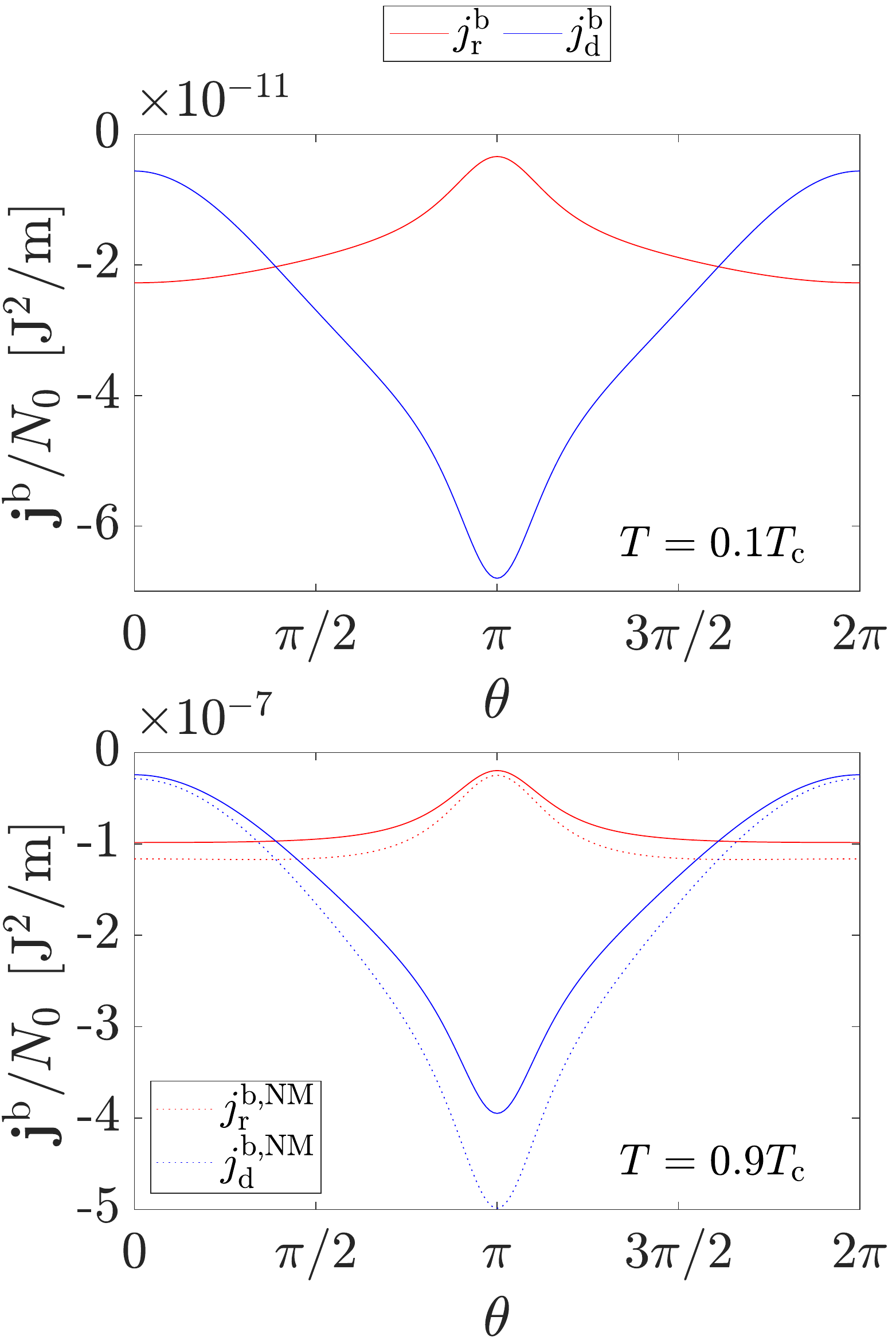}
\caption{The reactive (red) and dissipative (blue) backflow spin current, normalized to the density of states $N_0$, as function of $\theta$ through the left interface of a FMI|SC|FMI trilayer for two different temperatures, ${T = 0.1 T_{\rm c}}$ (upper plot) and ${T = 0.9 T_{\rm c}}$ (lower plot). We have used the parameters given in the main text, with $m_{\rm eff} = 0.1 \Delta_0$. In the lower plot, we have also plotted the spin current through an analogous FMI|NM|FMI trilayer (dotted lines).
}
\label{fig:SCandNMvsTheta}
\end{figure}

Fig.~\ref{fig:SCandNMvsTheta} plots the backflow spin current as function of $\theta$ for two different temperatures, ${T = 0.1 T_{\rm c}}$ and ${T = 0.9 T_{\rm c}}$. The spin pumping currents in both cases are purely reactive and equal to $j^{\rm p}/N_0 = 10^{-5}$ J$^2$/m. The first striking observation is that the spin current is much lower in the SC system at ${T = 0.1 T_{\rm c}}$ than at ${T = 0.9 T_{\rm c}}$. Singlet pair formation hinders injection of spin-currents into the superconductor. Next, we observe that the total spin current grows as $\theta$ approaches $\pi$, which is the case for both the SC and NM systems, and at both temperatures. This is due to the decreased impact of Hanle precession on the spin accumulation as the effective magnetic field decreases. Moreover, we note that the reactive spin current is favored close to $\theta = 0$, whereas the dissipative spin current is favored close to $\theta = \pi$. This is because the Hanle precession affects the reactive and dissipative spin current differently. Inspecting Eq.~\eqref{eq:analytical1}, we see that the reactive and dissipative spin current are suppressed by a factor $\propto (m_{\rm eff})^{-1}$ and $\propto (m_{\rm eff})^{-2}$ close to $\theta = 0$, respectively. For large effective magnetic fields, that is close to $\theta = 0$, the dissipative spin current is therefore strongly suppressed compared to the reactive spin current. Close to $\theta = \pi$, where Hanle precession is negligible, the reactive and dissipative spin current are suppressed $\propto V^{-2}$ and $\propto V^{-1}$, respectively, as can be seen in Eq. \eqref{eq:analytical2}. Hence, the dissipative spin current dominates close to $\theta = \pi$.
\begin{figure}[htb]
\centering
\includegraphics[width=1.0\columnwidth]{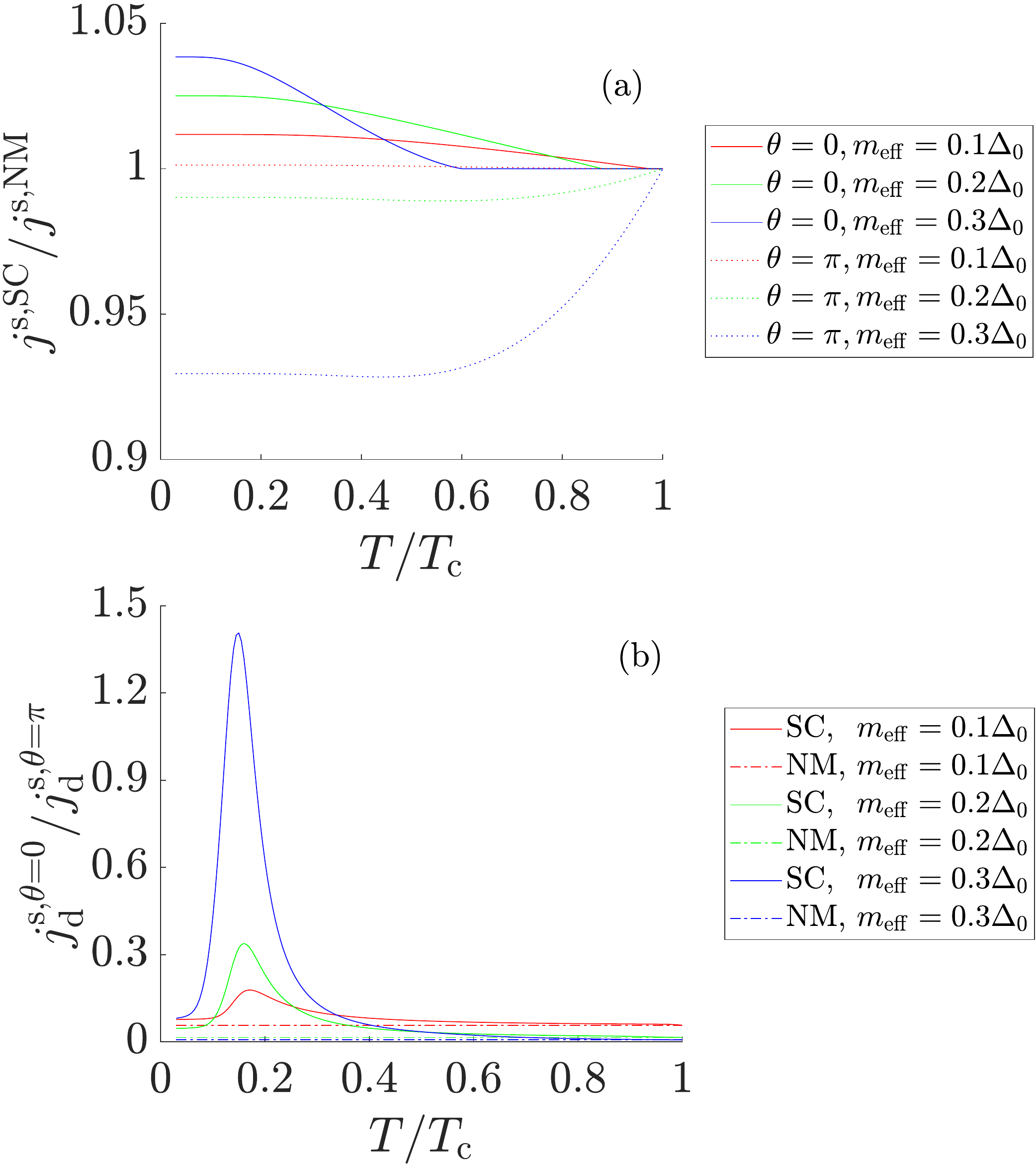}
\caption{(a) The temperature dependence of the total spin current in the superconducting system for two relative magnetization angles, $\theta = 0$ and $\theta = \pi$. The spin current is normalized to the normal metal limit, where the spin current is independent on temperature. (b) The ratio $j^{{\rm s}, \theta = 0}_{\rm d}/j^{{\rm s}, \theta = \pi}_{\rm d}$ plotted as function of temperature for both the SC and NM systems. We have used the parameters given in the main text. The lowest temperature included is $T = 0.03 T_{\rm c}$ in order to ensure that the gradient expansion is justified.}
\label{fig:SCandNMvsT}
\end{figure}

Let us now explore the temperature dependence in detail. In Fig. \ref{fig:SCandNMvsT}(a) we plot the total spin current as function of temperature for two angles, $\theta = 0$ and $\theta = \pi$, and for different effective field strengths $m_{\rm eff}$. We have normalized the spin currents with the respect to the analogous NM limit spin currents. The latter are independent on temperature. Due to the gradient expansion, the parameters must satisfy the condition $\beta^{-1} \gg \omega/\sqrt{8} \approx 0.003 k_{\rm B} T_{\rm c}$. We therefore restrict the temperature analysis to $T \geq 0.03 T_{\rm c}$. First, we observe that the spin currents approach the NM limit at the critical fields for the respective effective magnetic fields. We have already discussed this behavior, which is due to the amount quasiparticle states within a thermal energy interval from the Fermi energy. This entails an overall decrease in the total spin current for the $\theta = 0$ configuration, and an increase for the $\theta = \pi$ configuration. This is due to the nature of the backflow spin current. In the $\theta = 0$ configuration, the backflow spin current is dominated by a reactive component which counteracts the spin pumping current. In the $\theta = \pi$ configuration, the backflow spin current is dominated by a dissipative component. This spin current is oriented almost 90$^\circ$ relative to the spin pumping current, and therefore increases the total spin current.

Next, Fig.~\ref{fig:SCandNMvsT}(a) demonstrates that the temperature dependence of the normalized spin current for the $\theta = 0$ and $\theta = \pi$ states differ. In order to investigate this further, we plot the ratio between the dissipative spin currents in the parallel and anti-parallel configurations, $j^{{\rm s}, \theta = 0}_{\rm d}/j^{{\rm s}, \theta = \pi}_{\rm d}$, both in the NM and SC state, in Fig.~\ref{fig:SCandNMvsT}(b). Here, we observe that this ratio is a constant function of temperature in the NM limit, whereas it depends strongly on temperature in the superconducting state. The ratios peaks at slightly different temperatures for different effective fields $m_{\rm eff}$ in the superconducting state. The height of the peak increases with an increasing effective field $m_{\rm eff}$. As the temperature approaches $T_{\rm c}$, the ratio in the SC state converges towards the NM limit result.

This behavior is due to the aforementioned effect of temperature and effective magnetic field. In the parallel configuration, the effective magnetic fields of the two FMIs add constructively and cause a strong spin-splitting in the density of states. In the anti-parallel configuration, the effective fields add destructively and cause only a weakly spin-split density of states. At very low temperatures, the difference between the parallel and anti-parallel configuration is small for the chosen values of $m_{\rm eff}$. This is because neither state has a large density of states close to the almost $\delta$-function like thermal energy interval around the Fermi level. At slightly higher temperatures, the states which are pushed closer to the Fermi level starts overlapping with the thermal energy interval ${E_{\rm F} \pm \beta^{-1}}$. The difference between the two states is maximized for some intermediate temperature, ${k_{\rm B} T \lessapprox \Delta(T)}$, where we observe the peaks in Fig.~\ref{fig:SCandNMvsT}(b). At even higher temperatures, the thermal energy interval broadens further. The difference between the parallel and anti-parallel state then starts decreasing for higher temperatures, and eventually approaches the NM limit.

\subsection{Numerical analysis}
\label{ssec:num}

We here aim to briefly present a numerical solution to the problem which was solved analytically in Sec. \ref{ssec:lin1}. 
Our main goal is to evaluate whether the assumption of static gap can be justified to good approximation. A subsidiary goal is to show the time evolution of the gap, and the usefulness of a numerical method in this framework also when the static gap approximation is not valid.

We see from Eqs. \eqref{eq:Reffeq2} and \eqref{eq:Keffeq2} that the vectors $\ve{G}^{\rm A}$, $\ve{G}^{\rm R}$ and $\ve{G}^{\rm K}$ precess around the effective magnetic field. For such a class of equations, employing a fourth order Runge-Kutta method is suitable for obtaining a numerical solution. In order to test the validity of the static gap approximation, we want to perform a simulation of the system where the oscillations in the gap are maximized. This is expected to occur where the magnitude of the effective field oscillates with largest amplitude. From Eq. \eqref{eq:Delta1} one can show that this occurs at $\theta = \pi/2$ in the absence of relaxation, and we hence expect it to occur at $\theta = \pi/2$ also with the inclusion of relaxation.
\begin{figure}[htb]
\centering
\includegraphics[width=1.0\columnwidth]{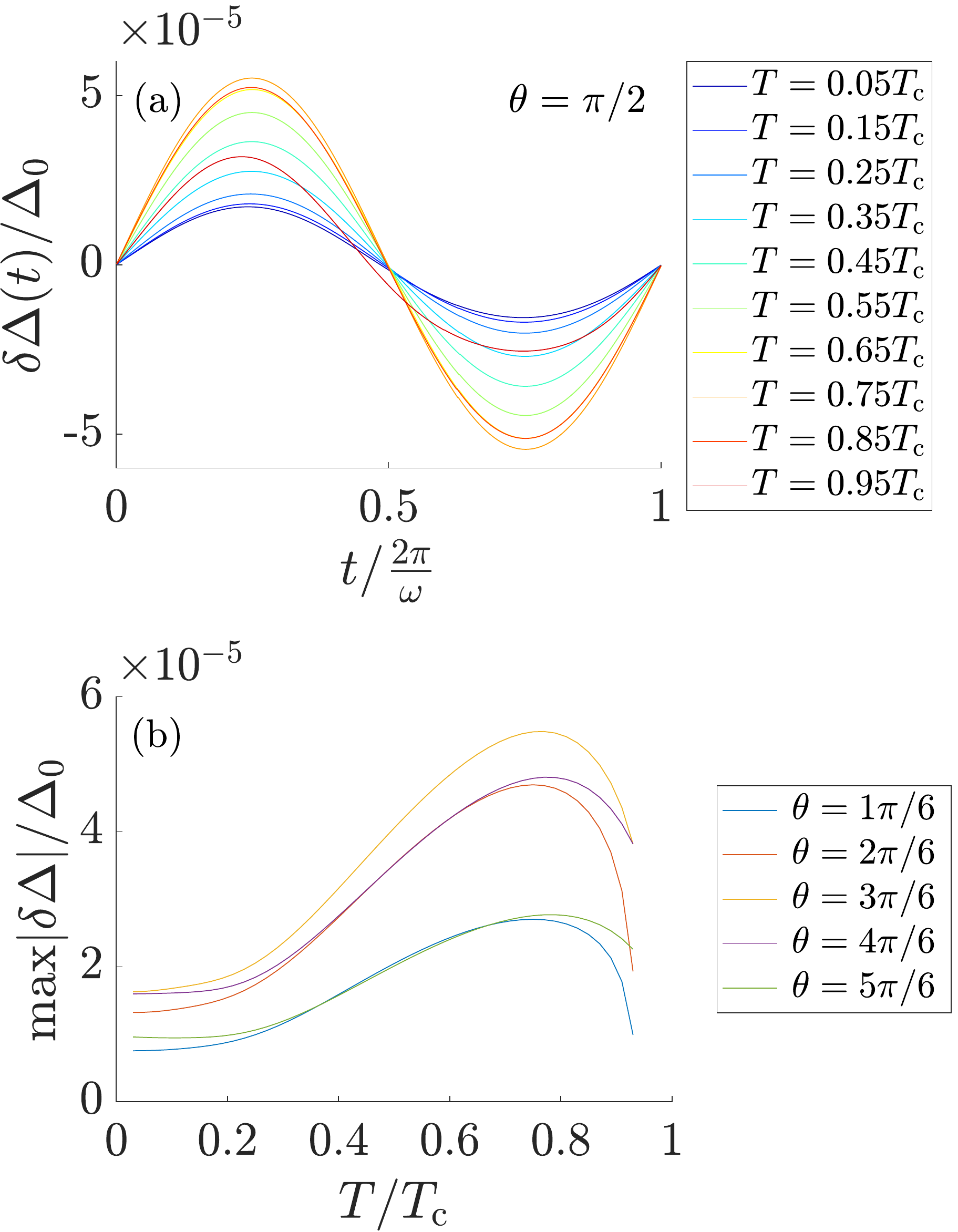}
\caption{(a) The fluctuations of the gap $\delta \Delta(t)$ plotted over one period $2 \pi/\omega$ at different temperatures for $\theta = \pi/2$. (b) The detailed temperature dependence of the gap fluctuation amplitude $\mathrm{max}|\delta\Delta|$ for different magnetization angles $\theta$. The gap fluctuations are normalized to the gap at zero temperature, $\Delta_0$, and we have used $m_{\rm eff} = 0.1 \Delta_0$.}
\label{fig:DeltavsTandTheta}
\end{figure}

Fig.~\ref{fig:DeltavsTandTheta}(a) for $\theta = \pi/2$ shows the fluctuation of the gap $\delta \Delta(t)$ normalized to $\Delta_0$ over one period $2\pi/\omega$ and for several temperatures $T$, with $m_{\rm eff} = 0.1 \Delta_0$. The gap oscillates harmonically with frequency $\omega$ for all temperatures up to $T = 0.85 T_{\rm c}$. At temperatures close the the critical temperature for the given effective magnetic field, the gap shows a non-linear response to the dynamical magnetization. This effect is visible for $T = 0.95 T_{\rm c}$, and is due to the increased sensitivity to fluctuations in the magnetic field as the temperature approaches the critical temperature. In Fig. \ref{fig:DeltavsTandTheta}(b), we further explore the $\theta$ and temperature dependence of the gap fluctuation amplitude, ${\mathrm{max}|\delta\Delta|}$, in the linear response regime. We observe that the fluctuations are largest at ${\theta = \pi/2}$, and that they are maximized at about $T \approx 0.8 T_{\rm c}$. Moreover, we observe that the fluctuations are not larger than about $5.5 \cdot 10^{-5} \Delta_0$. 
Let us now briefly remind the reader that the formal requirement for the static gap approximation was ${\delta \Delta \ll m_{\rm eff} \delta m}$, where $\delta m$ is the dynamic magnetization amplitude. We have ${m_{\rm eff} \delta m \approx 0.001 \Delta_0 \gg \delta \Delta \leq 5.5 \cdotp 10^{-5}} \Delta_0$, which implies that the static gap assumption is an excellent approximation in this instance.


\section{Conclusion}

We have derived an effective, time-dependent generalized Usadel equation in noncollinear FMI|SC|FMI trilayers with a thin superconducting layer and weakly magnetized FMIs. We have provided analytical solutions to these equations in terms of perturbations in the dynamic magnetization, first under the assumption of a static gap, and then a self-consistent solution in the absence of relaxation. Last, we have provided numerical procedures to obtain self-consistent solutions of the full equations without any further simplifications.

From the solutions to the generalized Usadel equation, we computed the spin currents generated by ferromagnetic resonance in one of the FMIs. We have explored this spin current as function of both temperature and relative magnetization angle between the FMIs. The spin current has been decomposed into a reactive and a dissipative part, which change the effective gyromagnetic ratio and Gilbert damping coefficient of the FMI. We found that the backflow spin current is generally largest when the magnetization orientations of the FMIs are anti-parallel. The ratio between the spin current in the parallel and anti-parallel configuration strongly depends on temperature in the SC. The origin is the Zeeman splitting of the quasiparticles at the gap edge. Last, we performed a numerical simulation to verify that the static gap assumption is a good approximation in our regime, also showing the usefulness of a numerical solution in this framework.
\\
\\
\textit{Acknowledgments.---}This work was supported by the Research Council of
Norway through its Centres of Excellence funding scheme,
Project No. 262633 "QuSpin", as well as by the European Research
Council via Advanced Grant No. 669442 "Insulatronics".

\appendix
\section{Additional parametrized Usadel equations}
\label{app:App1}

In the main text we provided four of the generalized Usadel equations, Eqs.~\eqref{eq:Reffeq1}-\eqref{eq:Keffeq2}, that were equations of motion for the normal components of the Green's functions. The remaining four equations that are needed to solve a system with non-zero anomalous Green's functions self-consistently are given as follows:

\begin{widetext}
\begin{align}
        F_0^{\rm R/A} &=  \frac{i\left(
        \Delta \left( m_{\rm eff} \ve{m} \cdot \ve{G}^{\rm R/A} - (E - i V) G_0^{\rm R/A} \right)
        - \frac{i m_{\rm eff}^2}{2} \ve{m} \cdot \left(\frac{\partial \ve{m}}{\partial t}  \times \frac{\partial \ve{F}^{\rm R/A}}{\partial E}\right)  \right)}{m_{\rm eff}^2 \ve{m}^2 - (E - i V)^2} , \label{eq:AppReffeq3}\\
        \ve{F}^{\rm R/A} &= \frac{i \Delta \ve{G}^{\rm R/A}}{(E - i V)} - \frac{i\left( 
        m_{\rm eff} \Delta \left( m_{\rm eff} \ve{m}\cdot\ve{G}^{\rm R/A} - (E - i V) G_0^{\rm R/A} \right) \ve{m} 
        - \frac{ i m_{\rm eff}^3}{2} \left[ \ve{m} \cdot \left( \frac{\partial \ve{m}}{\partial t} \times \frac{\partial \ve{F}^{\rm R/A}}{\partial E} \right) \right] \ve{m} 
        + \frac{i m_{\rm eff}}{2} (m_{\rm eff}^2 \ve{m}^2 - (E - i V)^2) \left( \frac{\partial \ve{m}}{\partial t} \times \frac{\partial \ve{F}^{\rm R/A}}{\partial E} \right) 
        \right)}{(E - i V)(m_{\rm eff}^2 \ve{m}^2 - (E - i V)^2)}
        \label{eq:AppReffeq4} \\
    \begin{split}
        F_0^{\rm K} &=  \frac{i\left(
        \Delta \left( m_{\rm eff} \ve{m} \cdot \ve{G}^{\rm K} - E G_0^{\rm K} \right)
        - \frac{i m_{\rm eff}^2}{2} \ve{m} \cdot \left(\frac{\partial \ve{m}}{\partial t}  \times \frac{\partial \ve{F}^{\rm K}}{\partial E}\right)  \right)}{m_{\rm eff}^2 \ve{m}^2 - E^2} 
        - \frac{i V \tanh\left(\frac{\beta E}{2}\right) \left[ \ve{m} \cdot \left( \ve{F}^{\rm R} + \ve{F}^{\rm A} \right) - E \left( F_0^{\rm R} + F_0^{\rm A} \right) \right]}{\left( m_{\rm eff}^2 \ve{m}^2 - E^2 \right)} \\
        &\quad - \frac{\beta  V \sech^2\left(\frac{\beta E}{2}\right) \left[ \ve{m} \cdot \left( \frac{\partial\ve{F}^{\rm R}}{\partial t} - \frac{\partial\ve{F}^{\rm A}}{\partial t} \right) - E \left( \frac{\partial F_0^{\rm R}}{\partial t} - \frac{\partial F_0^{\rm A}}{\partial t} \right) \right]}{4\left( m_{\rm eff}^2 \ve{m}^2 - E^2 \right)},
    \end{split} \label{eq:AppKeffeq3}\\
    \begin{split}
        \ve{F}^{\rm K} &= \frac{i \Delta \ve{G}^K}{E} - \frac{i\left( 
        m_{\rm eff} \Delta \left( m_{\rm eff} \ve{m}\cdot\ve{G}^{\rm K} - E G_0^{\rm K} \right) \ve{m} 
        - \frac{ i m_{\rm eff}^3}{2} \left[ \ve{m} \cdot \left( \frac{\partial \ve{m}}{\partial t} \times \frac{\partial \ve{F}^{\rm K}}{\partial E} \right) \right] \ve{m} 
        + \frac{i m_{\rm eff}}{2} (m_{\rm eff}^2 \ve{m}^2 - E^2) \left( \frac{\partial \ve{m}}{\partial t} \times \frac{\partial \ve{F}^{\rm K}}{\partial E} \right) 
        \right)}{E(m_{\rm eff}^2 \ve{m}^2 - E^2)} \\
        &\quad+ \frac{i V \tanh\left(\frac{\beta E}{2}\right) \left[ \left\{\ve{m} \cdot \left( \ve{F}^{\rm R} + \ve{F}^{\rm A} \right) - E \left( F_0^{\rm R} + F_0^{\rm A} \right) \right\} \ve{m} - \left( \ve{m}^2 - E^2 \right) \left( \ve{F}^{\rm R} + \ve{F}^{\rm A} \right) \right]}{E \left( m_{\rm eff}^2 \ve{m}^2 - E^2 \right)} \\
        &\quad + \frac{\beta  V \sech^2\left(\frac{\beta E}{2}\right) \left[ \left\{ \ve{m} \cdot \left( \frac{\partial\ve{F}^{\rm R}}{\partial t} - \frac{\partial\ve{F}^{\rm A}}{\partial t} \right) - E \left( \frac{\partial F_0^{\rm R}}{\partial t} - \frac{\partial F_0^{\rm A}}{\partial t} \right) \right\} \ve{m} - \left( \ve{m}^2 - E^2 \right) \left( \frac{\partial\ve{F}^{\rm R}}{\partial t} - \frac{\partial\ve{F}^{\rm A}}{\partial t} \right) \right]}{4 E \left( m_{\rm eff}^2 \ve{m}^2 - E^2 \right)}
        ,
    \end{split}\label{eq:AppKeffeq4}
\end{align}
\end{widetext}
\noindent where the notation is defined in the main text.

\section{Self-consistent solution in the absence of spin relaxation}
\label{ssec:lin2}
We will here derive a self-consistent solution to the generalized Usadel Equations, Eqs.~\eqref{eq:Reffeq1}-\eqref{eq:Keffeq2}, Eqs.~\eqref{eq:AppReffeq3}-\eqref{eq:AppKeffeq4}, and the gap equation \eqref{eq:gap}, in the absence of spin relaxation ($V = 0$). This solution has restricted physical relevance, and only applies in the limit where the precession frequency is much larger than the relaxation rate. However, it is included as a proof of concept that a self-consistent solution is in principle possible.

The derivation follows the lines of what was presented in Sec. \ref{ssec:lin1}, with a few exceptions. In addition to the perturbation expansion in Eqs. \eqref{eq:pert1}, we also expand
\begin{align}
\begin{split}
     F_0 &= F_0^{(0)} + F_0^{(1)} + F_0^{(2)} + (...), \\
     \ve{F} &= \ve{F}^{(0)} + \ve{F}^{(1)} + \ve{F}^{(2)} + (...) ,\\
     \Delta &= \Delta^{(0)} + \Delta^{(1)} + \Delta^{(2)} + (...).
 \end{split}\label{eq:pert2}
\end{align}

\noindent We have dropped the retarded/advanced and Keldysh superscript in order to keep the derivation as general as possible. This derivation hence applies to all Green's function components. We also propose one additional ansatz,
\begin{align}
\begin{split}
    \Delta^{(1)} &= \Delta^{(1)}_{+} e^{i \omega t} + \Delta^{(1)}_{-} e^{-i \omega t}.
\end{split}\label{eq:ans2}
\end{align}
If we insert this into the generalized Usadel equations to first order in $\delta m$, and with $V = 0$, we obtain the solutions
\begin{align}
    G_{0 \pm}^{(1)} &=  m_{\rm eff} \left(\frac{\partial \ve{G}^{(0)}}{\partial E}\right) \cdot \ve{m}_{\pm}^{(1)} -i \left(\frac{\partial F_0^{(0)}}{\partial E}\right) \Delta^{(1)}_{\pm}  , \label{eq:G00} \\
    \ve{G}_{\pm}^{(1)} &= m_{\rm eff} \tilde{\mathbf{A}}_{\pm\omega}^{-1}\tilde{\mathbf{B}}_{\pm\omega} \ve{m}^{(1)}_{\pm} \pm\omega \Delta^{(1)}_{\pm} \tilde{\mathbf{A}}_{\pm\omega}^{-1} \frac{\partial \ve{F}^{(0)}}{\partial E} \label{eq:G0} ,
\end{align}

\noindent where

\begin{align}
    \begin{split}
        \tilde{\mathbf{A}}_{\pm\omega} =
        \begin{pmatrix}
            \pm i \omega & 2 m_{\rm eff} -m_z^{(0)} & 2 m_{\rm eff} m_y^{(0)} \\
            2 m_{\rm eff} m_z^{(0)} & \pm i \omega & -2 m_{\rm eff} m_x^{(0)} \\
            -2 m_{\rm eff} m_y^{(0)} & 2 m_{\rm eff} m_x^{(0)} & \pm i \omega
        \end{pmatrix}
    \end{split}
\end{align}

\noindent and

\begin{align}
    \begin{split}
        \tilde{\mathbf{B}}_{\pm\omega} =
        \begin{pmatrix}
            \pm i \omega \frac{\partial G_0^{(0)}}{\partial E}  & -2 G_z^{(0)} & 2 G_y^{(0)} \\
            2 G_z^{(0)} & \pm i \omega \frac{\partial G_0^{(0)}}{\partial E}  & -2 G_x^{(0)} \\
            -2 G_y^{(0)} & 2 G_x^{(0)} & \pm i \omega \frac{\partial G_0^{(0)}}{\partial E}
        \end{pmatrix}.
    \end{split}
\end{align}

To solve for $\Delta^{(1)}(t)$, we look closer at the gap equation given in Eq.~\eqref{eq:gap}. If we insert the generalized Usadel equation for $F_0^{\rm K}$ [Eq.~\eqref{eq:AppKeffeq3}] into the gap equation while using $V = 0$, divide both sides by $\Delta$, and assume that $|m_{\rm eff} \ve{m}^{(1)}| \ll \delta$, the first and second order gap equations read
\begin{align}
    1 =& \frac{N_0 \lambda}{4} \int_{-\omega_D}^{\omega_D} \mathrm{d} E \frac{\left( m_{\rm eff} \ve{m}^{(0)} \cdot \ve{G}^{\mathrm{K} (0)} - E G_0^{\mathrm{K} (0)} \right)}{m_{\rm eff}^2 (\ve{m}^{(0)})^2 - E^2}, \label{eq:gap1}\\
    \begin{split}
        0 =& \int_{-\omega_D}^{\omega_D} \mathrm{d} E \frac{1}{m_{\rm eff}^2 (\ve{m}^{(0)})^2 - E^2} \\
        &\Bigg\{
        m_{\rm eff}\left( \ve{m}^{(1)} \cdot \ve{G}^{\mathrm{K} (0)} + \ve{m}^{(0)} \cdot \ve{G}^{\mathrm{K} (1)} \right) 
         - E G_0^{\mathrm{K} (1)} \\
        &- 
        2 m_{\rm eff}^2 \left( \ve{m}^{(0)} \cdot \ve{m}^{(1)} \right) \frac{\left( m_{\rm eff} \ve{m}^{(0)} \cdot \ve{G}^{\mathrm{K} (0)} - E G_0^{\mathrm{K} (0)} \right)}{m_{\rm eff}^2 (\ve{m}^{(0)})^2 - E^2}
        \Bigg\}.
    \end{split} \label{eq:gap2}
\end{align}

\noindent Here, we used ${\ve{m}^{(0)} \cdot \left( \frac{\partial \ve{m}^{(1)}}{\partial t} \times \frac{\partial \ve{F}^{\mathrm{K} (0)}}{\partial E} \right) = 0}$, since ${\ve{F}^{\mathrm{K} (0)} \parallel \ve{m}^{(0)}}$. We have moreover used ${E \rightarrow E + i \delta}$, and ${m_{\rm eff} |\ve{m}^{(0)} \cdot \ve{m}^{(1)}| \ll  |\ve{m}^{(0)}| \delta}$, ensuring that the expansion is also valid when $\re{E} \rightarrow m_{\rm eff} |\ve{m}^{(0)}|$. Eq.~\eqref{eq:gap1} is simply the zeroth order gap equation, while Eq.~\eqref{eq:gap2} must be used to find self-consistent solution to the first order Green's function components. All that remains now is to insert the ansätze Eqs.~\eqref{eq:ans1} and \eqref{eq:ans2} into Eq.~\eqref{eq:gap2}. The resulting solution for the first order components $\Delta^{(1)}_\pm$ reads
\begin{align}
\begin{split}
    \Delta^{(1)}_{\pm} &= \frac{1}{T_{\pm}}\int_{-\omega_D}^{\omega_D} \mathrm{d} E \frac{m_{\rm eff}}{m_{\rm eff}^2 (\ve{m}^{(0)})^2 - E^2} \Bigg\{ 
    \left(\ve{m}^{(1)}_{\pm}\cdot\ve{G}^{\mathrm{K} (0)}\right)\\ 
    &+ m_{\rm eff} \left(\left[\mathbf{A}_{\pm\omega}^{-1}\mathbf{B}_{\pm\omega}\ve{m}^{(1)}_{\pm}\right]\cdot \ve{m}^{(0)}\right) 
    - E \left(\ve{m}^{(1)}_{\pm} \cdot \frac{\partial \ve{G}^{\mathrm{K} (0)}}{\partial E} \right)\\
    &- 2 m_{\rm eff} \left(\ve{m}^{(1)}_{\pm} \cdot \ve{m}^{(0)} \right) \frac{\left( m_{\rm eff} \ve{m}^{(0)} \cdot \ve{G}^{\mathrm{K} (0)} - E G_0^{\mathrm{K} (0)} \right)}{m_{\rm eff}^2 (\ve{m}^{(0)})^2 - E^2} 
    \Bigg\},
\end{split}
\label{eq:Delta1}
\end{align}

\noindent where $T_\pm$ is defined by
\begin{align}
\begin{split}
    T_{\pm} &= -\int_{-\omega_D}^{\omega_D} \mathrm{d} E \frac{\pm \omega m_{\rm eff} \left[ \mathbf{A}_{\pm\omega}^{-1} \frac{\partial \ve{F}^{\mathrm{K} (0)}}{\partial E}\right]\cdot\ve{m}^{(0)} + i E \frac{\partial F^{\mathrm{K} (0)}_0}{\partial E}}{m_{\rm eff}^2 (\ve{m}^{(0)})^2 - E^2}.
\end{split}
\end{align}


\renewcommand{\refname}{Bibliography}
%

\end{document}